\documentclass[aps,pre,onecolumn]{revtex4}
\usepackage{graphicx, bm, bbm,amsmath, amssymb, epsfig,color}
\usepackage{todonotes}

\pdfoutput=1

\newcommand{\nn}{\noindent}

\newcommand{\bq}{\begin{align}}
\newcommand{\eq}{\end{align}}

\usepackage{caption}
\usepackage{subcaption}
\begin{document}
\title{Efficient sliding locomotion of three-link bodies with inertia}

\author{Adam Earnst}
\author{Silas Alben}
\email{alben@umich.edu}
\affiliation{Department of Mathematics, University of Michigan, Ann Arbor, MI 48109, USA}

\date{\today}

\begin{abstract}
Many previous studies of sliding locomotion have assumed that body inertia is negligible. Here we optimize the kinematics of a three-link body for efficient locomotion and include among the kinematic parameters the temporal period of locomotion, or equivalently, the body inertia.    
The optimal inertia is non-negligible when the coefficient of friction for sliding transverse to the body axis is small. Inertia is also significant in a few cases with relatively large coefficients of friction for transverse and backward sliding, and here the optimal motions are less sensitive to the inertia parameter. The optimal motions seem to converge as the number of frequencies used is increased from one to four. For some of the optimal motions with significant inertia we find dramatic reductions in efficiency when the inertia parameter is decreased to zero. For the motions that are optimal with zero inertia, the efficiency decreases more gradually when we raise the inertia to moderate and large values.
\end{abstract}

\maketitle
\section{Introduction}

This work addresses the physics of terrestrial locomotion, part of a larger field of interdisciplinary studies of the locomotion of organisms, robots, and vehicles, often bio-inspired \cite{bekker1956theory,purcell1977life,hirosebiologically,dickinson2000animals,choset2005principles,aguilar2016review,biewener2018animal}. Aerial and aquatic locomotion problems are often dominated by the interaction between a locomoting body (or bodies) and the surrounding fluid, and may involve complicated fluid dynamics \cite{dickinson2000animals,biewener2018animal}. Terrestrial locomotion is usually dominated by local contact forces involving friction \cite{hirosebiologically,choset2005principles,Maladen:2009es,maladen2011undulatory,aguilar2016review}, which may also be complicated to characterize. Here we study sliding locomotion inspired by biological and robotic snakes \cite{gray1946mechanism,jayne1986kinematics,hirosebiologically,choset2005principles,lillywhite2014snakes,aguilar2016review}. As with several recent models \cite{ma2001analysis,sato2002serpentine,chernousko2005modelling,GuMa2008a,HuNiScSh2009a,HuSh2012a,aguilar2016review,yona2019wheeled,zhang2021friction,rieser2021functional}, we use a local Coulomb friction force model for the interaction between the body and the surface that it slides across. The problem is similar to a larger body of work that has used resistive force theory to approximate fluid forces on swimming bodies in the viscous-dominated (zero-Reynolds-number or Stokes) regime \cite{purcell1977life,taylor1952analysis,childress1981mechanics,lauga2009hydrodynamics,HaBuHoCh2011a}. For the sliding locomotion problem here we adopt the three-link body that has been used Purcell and many others; it is one of the simplest bodies that can locomote at zero Reynolds number, by performing time-irreversible (e.g. undulatory) motions \cite{purcell1977life,childress1981mechanics,BeKoSt2003a,TaHo2007a,AvRa2008a}. Unlike the Stokes swimmers, for sliding locomotion the body's inertia can play an important role, though it has mostly been neglected for simplicity \cite{GuMa2008a,HuNiScSh2009a,HuSh2012a,peng2016characteristics}. Neglecting inertia allows one to analyze this system and the three-link swimmer using a geometrical-mechanics framework \cite{alouges2013self,hatton2017kinematic,bettiol2017purcell,bittner2018geometrically,yona2019wheeled}.

The body's inertia can be neglected for motions with small accelerations, which is a reasonable approximation for some but not all biological snake motions \cite{HuNiScSh2009a,lillywhite2014snakes}. Important exceptions include predator-prey interactions and fast steady locomotion \cite{greenwald1978kinematics,alfaro2003sweeping,gans1984slide,lillywhite2014snakes,wang2018dynamics}. 
The main subject of the present work is the effect of body inertia on sliding locomotion (biological or robotic).
Can such motions be more efficient than those with negligible inertia? We will show that in some cases the answer is yes. We will study three-link bodies because they are simpler to explore and optimize than general smooth bodies yet can still approximate familiar snake motions such as lateral undulation. In the simpler case of a fore-aft symmetric two-link body, inertia is necessary for locomotion, and such locomotion can be relatively efficient, in the limit of small friction normal to the links \cite{alben2020intermittent}. In this limit the peak efficiency of a two-link body can be slightly over half the peak for all planar sliding bodies and motions \cite{alben2020intermittent}.    

Unlike for a two-link body, inertia is not necessary for the sliding locomotion of a three-link body. In some locomotion problems there is a trade-off between speed and efficiency, where the highest efficiency is achieved at the lowest speeds. This was found by \cite{ariizumi2017dynamic} in their model of snake locomotion using sidewinding, lateral undulation, and
sinus lifting gaits, and by \cite{saito2002modeling,baysal2020optimally} in models of lateral undulation at particular sets of friction
coefficients. Here we compute optimally efficient gaits and find that the most efficient gait occurs with a small velocity in only a portion of friction coefficient space. In some regions the optimally efficient gait has an $O(1)$ speed, so there slower speeds require a decrease in efficiency.

This work builds upon a few previous works that used the same Coulomb friction model for the forces on a planar body sliding on a surface. Different specialized methods were used to compute the body's dynamics and perform optimization in different cases.
Motions of bodies {\it without} inertia---two-link bodies with a wide range of friction coefficients and three-link bodies with a specific biologically motivated set of friction coefficients---were studied in \cite{JiAl2013}, using asymptotic analysis and computations, including a commercial optimization toolbox. Optimal motions of smooth bodies were found using quasi-Newton methods and asymptotic analysis in \cite{AlbenSnake2013,wang2014optimizing}. Motions of smooth, passively flexible bodies with inertia were simulated using a Broyden iterative method \cite{wang2018dynamics}.
The optimally efficient motions of a three-link body {\it without} inertia were studied in \cite{alben2019efficient} with isotropic friction and
in \cite{alben2021efficient} with anisotropic friction. In the absence of body inertia, the invariance of the dynamics under time reparametrization allowed the motions of three-link bodies to be solved using a look-up table. $O(10^6)$ motions were simulated, providing a view of performance across kinematic space, at many friction coefficient values. 
Optimally efficient motions of a two-link body {\it with} inertia were studied in \cite{alben2020intermittent}, using explicit time-stepping and Newton's method for optimization. The present work studies the optimal motions of a three-link body {\it with} inertia. With three links, is not easy to formulate an explicit time-stepping scheme for the dynamics, so we use an implicit Newton-based solver at each time step. Like \cite{alben2021efficient}, we use a population-based optimization method. Three-dimensional (nonplanar) motions \cite{marvi2014sidewinding,zhang2021friction,alben2022efficient} will not be considered here for simplicity.


We find that optimal efficiency occurs with negligible inertia in one contiguous region of friction parameter space, where friction normal to the links is moderate to high. Optimal efficiency occurs with nonzero and moderate inertia when normal friction is low, and surprisingly, also occurs with moderate inertia where normal friction is very high, and with large inertia where backward friction is very high. The rest of the paper is as follows: section \ref{sec:Model} presents the dynamical model, and sections \ref{sec:ellipse} and
\ref{sec:multiple} present the optimal motions in the single- and multiple-frequency cases, respectively. Section \ref{sec:summary} summarizes the findings.

\section{Model \label{sec:Model}}

\begin{figure} [h]
           \begin{center}
           \begin{tabular}{c}
               \includegraphics[width=6.4in]{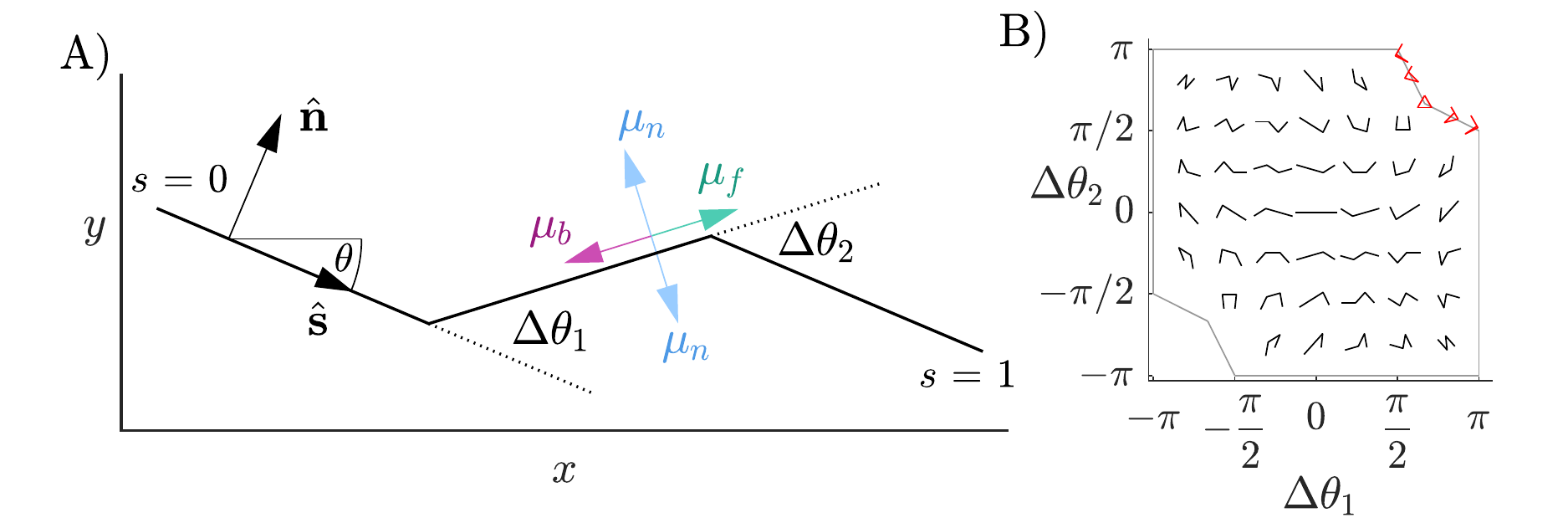} \\
           \vspace{-.25in}
           \end{tabular}
          \caption{\footnotesize A) Schematic diagram of a 
three-link body with changes in angles $\Delta\theta_1$ (here positive) and
$\Delta\theta_2$ (here negative) between the links. The body is
parametrized by arc length $s$ (nondimensionalized by body length), at an instant in time. The tangent angle $\theta$
and the unit vectors
tangent and normal to the body at a point, $\hat{\mathbf{s}}$ and $\hat{\mathbf{n}}$ respectively, are labeled. Vectors representing
forward, backward, and normal velocities are shown with
the corresponding friction coefficients $\mu_f$, $\mu_b$, and $\mu_n$.
B) Examples of body shapes in the ($\Delta\theta_1$, $\Delta\theta_2$)-plane.
Shapes that do not self-intersect are shown in black; a few shapes at the threshold of
self-intersection are shown in red. Figure
reproduced from \cite{alben2021efficient}.
 \label{fig:ThreeLinkSchematic}}
           \end{center}
         \vspace{-.10in}
        \end{figure}

Our model is essentially the same as in \cite{alben2021efficient} but with the addition of body inertia. We give the main points here for completeness.
The model describes the dynamics of a shape-changing body acted on by Coulomb friction with the ground, as in \cite{HuNiScSh2009a,HuSh2012a,JiAl2013} and other
recent studies. 
We represent the body as a polygonal curve $\mathbf{X}(s,t) = (x(s,t), y(s,t))$,
parametrized by arc length $s$ and varying with time $t$. A schematic
diagram is shown in figure \ref{fig:ThreeLinkSchematic}A. 

The basic problem is to prescribe the time-dependent
shape of the body in order to obtain efficient locomotion. The shape is
described by $\Delta\theta_1(t)$ and $\Delta\theta_2(t)$, the differences
between the tangent angles of the adjacent links. 
A representative set of body shapes is plotted at the corresponding ($\Delta\theta_1$,$\Delta\theta_2$) locations
in figure \ref{fig:ThreeLinkSchematic}B. The region inside the gray polygonal boundary consists of shapes that do not self-intersect. Five examples of shapes that lie on the boundary are shown in red (along the upper right portion of the boundary). In this work we will consider time-periodic kinematics, which are represented by closed curves in the
($\Delta\theta_1$,$\Delta\theta_2$)-plane.

To write the dynamical equations (Newton's laws), we first write
the body tangent angle as $\theta(s,t)$; it satisfies 
$\partial_s x = \cos\theta$ and $\partial_s y = \sin\theta$.
The unit vectors
tangent and normal to the body are $\hat{\mathbf{s}} = (\partial_s x, \partial_s y)$ 
and $\hat{\mathbf{n}} = (-\partial_s y, \partial_s x)$
respectively. We write 
\begin{align}
\theta(s,t) &= \theta_0(t) + \Delta \theta_1(t) H(s-1/3) + \Delta \theta_2(t) H(s-2/3) \label{theta0}
\end{align}
\nn where $H$ is the Heaviside function and $\theta_0(t)$ is the tangent
angle at the ``tail" (the $s=0$ end), an unknown to be solved for using Newton's equations of motion.
The body position is obtained by integrating the tangent vector:
\begin{align}
x(s,t) & = x_0(t) + \int_0^s \cos \theta(s',t) ds', \label{x0}\\
y(s,t) &= y_0(t) + \int_0^s \sin \theta(s',t) ds'. \label{y0}
\end{align}
\nn The tail position $\mathbf{X}_0(t) = (x_0(t), y_0(t))$ and tangent angle $\theta_0(t)$
are determined by the force and torque
balance for the body, i.e. Newton's second law \cite{HuNiScSh2009a,HuSh2012a}:
\begin{align}
\int_0^L \rho \partial_{tt} x ds &= \int_0^L f_x ds, \label{fx0} \\
\int_0^L \rho \partial_{tt} y ds &= \int_0^L f_y ds, \label{fy0} \\
\int_0^L  \rho \mathbf{X}^\perp \cdot \partial_{tt} \mathbf{X} ds
&= \int_0^L \mathbf{X}^\perp \cdot \mathbf{f} ds. \label{torque0}
\end{align}
\nn Here  $L$ is the body length, $\rho$ is the body's mass per unit length, and
$\mathbf{X}^\perp = (-y,x)$. For simplicity, the body is assumed to be locally inextensible
so $L$ is constant in time.
$\mathbf{f}$ is the force per unit length on the body
due to Coulomb friction with the ground:
\begin{align}
\mathbf{f}(s,t) &\equiv -\rho g\mu_n
\left( \widehat{\partial_t{\mathbf{X}}}_\delta\cdot \hat{\mathbf{n}} \right)\hat{\mathbf{n}}
- \rho g\left( \mu_f H(\widehat{\partial_t{\mathbf{X}}}_\delta\cdot \hat{\mathbf{s}})
+ \mu_b (1-H(\widehat{\partial_t{\mathbf{X}}}_\delta\cdot \hat{\mathbf{s}}))\right)
\left( \widehat{\partial_t{\mathbf{X}}}_\delta\cdot \hat{\mathbf{s}} \right)\hat{\mathbf{s}}, \label{frictiondelta} \\ 
\widehat{\partial_t{\mathbf{X}}}_\delta &\equiv \frac{\left(\partial_t x, \partial_t y\right)}{\sqrt{\partial_t x^2 +\partial_t y^2 + \delta^2}}, \label{delta}
\end{align}
\nn \textcolor{black}{and $g$ is gravitational acceleration.}
Again $H$ is the Heaviside function, and $\widehat{\partial_t{\mathbf{X}}}_\delta$ is the normalized velocity,
regularized with a small parameter $\delta = 10^{-2}$ here. Nonzero $\delta$ aids the numerical solutions, particularly when $\rho$ is small, but $\delta$ has little effect on the solutions as long as it is much smaller than the scale of body velocities (typically $O$(1)), as noted in \cite{alben2019efficient} and \cite{alben2021efficient} in the case of zero inertia. We find empirically that there is little change in the results (less than 1\% in relative magnitude) when $\delta$ is decreased below $10^{-2}$.

According to (\ref{frictiondelta}) the body experiences
friction with coefficients $\mu_f$, $\mu_b$, and $\mu_n$ for motions
in the forward ($\hat{\mathbf{s}}$), backward ($-\hat{\mathbf{s}}$),
and normal ($\pm\hat{\mathbf{n}}$) directions, respectively. 
If $\mu_b \neq \mu_f$, we define the forward direction so that
$\mu_f < \mu_b$, without loss of generality.  
In general the body velocity at a given point has both tangential and
normal components, and the frictional force density has
components acting in each direction. A similar decomposition of force
into directional components
occurs for viscous fluid forces on slender bodies \cite{cox1970motion}. 

We nondimensionalize equations (\ref{fx0})--(\ref{torque0}) by dividing
lengths by the body length $L$ and mass by $\rho L$.
We nondimensionalize the time $t$ by $\sqrt{L/g}$, and assume that
the body shape $(\Delta\theta_1(t), \Delta\theta_2(t))$ is periodic in time (as is typical for steady locomotion \cite{HuNiScSh2009a}) with dimensionless period $T$, a parameter that we are free to choose. We compute the motions with respect to 
a time variable $\tau = t/T$,
the time scaled by the period, so that a population of motions with different periods can be 
computed as an ensemble on the same $\tau$ grid. Using $\tau$, the dimensionless equations
(\ref{fx0})--(\ref{torque0}) are
\begin{align}
\frac{1}{T^2} \int_0^1 \partial_{\tau\tau} x ds &= \int_0^1 f_x ds, \label{fxa} \\
\frac{1}{T^2}\int_0^1 \partial_{\tau\tau} y ds &= \int_0^1 f_y ds, \label{fya} \\
\frac{1}{T^2}\int_0^1 \mathbf{X}^\perp \cdot \partial_{\tau\tau} \mathbf{X} ds
&= \int_0^1 \mathbf{X}^\perp \cdot \mathbf{f} ds. \label{torquea}
\end{align}
\nn In (\ref{fxa})--(\ref{torquea}) and from now on, all variables are
dimensionless. If we divide by $\mu_f$ in each of (\ref{fxa})--(\ref{torquea}), we can see that the problem depends on 
three parameters: $R \equiv 1/\mu_f T^2$, $\mu_b/\mu_f$, and $\mu_n/\mu_f$. 

In most of the previous studies using this model (e.g. \cite{HuNiScSh2009a,HuSh2012a,AlbenSnake2013, alben2021efficient}),
$T$ is taken to infinity, so the left sides of (\ref{fxa})--(\ref{torquea}) vanish. Here we allow $T$ to be finite, and it (actually, $R$) is one of the parameters we optimize with respect to.
Physically, $T$ or $R$ are only rough measures of the sizes of the inertia terms, which are actually the full left sides of (\ref{fxa})--(\ref{torquea}), and therefore depend on the body kinematics.
For example, for lateral undulation, given by a periodic traveling wave of deflection of a nearly straight body (considering for the moment a smooth body rather than a three-link body), the integrals on the left sides of
(\ref{fxa})--(\ref{torquea}) are small even if $T = O(1)$, if the body length is a large integer multiple of the spatial period of deflection. In such a case (studied in \cite{AlbenSnake2013,alben2019efficient}), $T$ could be small (or $R$ large) but the inertia terms are small. Such a motion and its efficiency are insensitive to the value of $R$. For the optimal solutions of the three-link body studied here, we find wide variations in the sensitivity of efficiency to $R$. Some optimal motions work well across a wide range of $R$ and others do not. 

One can also think of $T$ or $R$ as measures of the speed of the motion, although the true speed of the motion is the ratio of the center-of-mass displacement per period to $T$. The center-of-mass displacement per period is generally not larger than the body length, so in general a large $T$ implies a small locomotion speed. However, a small $T$ could occur with small or large speeds, depending on the center-of-mass displacement.
A more straightforward but less physically meaningful interpretation of $R$ is as simply one of the kinematic parameters that can be tuned to change the body's motion and its efficiency.


When $(\Delta\theta_1(\tau), \Delta\theta_2(\tau))$ is periodic, the body motion may be solved as
an initial value problem by giving initial conditions for $\{x_0(\tau), y_0(\tau), \theta_0(\tau),
\dot{x}_0(\tau), \dot{y}_0(\tau), \dot{\theta}_0(\tau)\}$ (all set to zero here), 
and then evolving them forward in time using
(\ref{fxa})--(\ref{torquea}). We find that generically the resulting body motion $\mathbf{X}(s,\tau)$ evolves 
towards a time-periodic state with the same $\tau$ period (unity) as $(\Delta\theta_1(\tau), \Delta\theta_2(\tau))$.
Part of our definition of locomotor efficiency is the body's average speed---i.e. the magnitude of the
average velocity of the body's center of mass---over a finite length of time $T_1$ starting at time $t_0$:
\begin{align}
\| \langle \overline{\partial_t \mathbf{X}} \rangle \|
&\equiv \Biggl \| \frac{1}{T_1}\int_{t_0}^{t_0 + T_1} \int_0^1 \partial_t \mathbf{X} \,ds \,dt \Biggr \| \\
&= \frac{1}{T_1}\sqrt{ \left( \int_0^1 x(s,t_0+T_1) - x(s,t_0) ds\right)^2 +
\left(\int_0^1 y(s,t_0+T_1) - y(s,t_0) ds\right)^2} \\ 
&= \frac{1}{T_1} \sqrt{ \left( \int_{0}^{1} x(s,\tau_0+\mathcal{T}_1) - x(s,\tau_0) ds\right)^2 +
\left(\int_{0}^{1} y(s,\tau_0+\mathcal{T}_1) - y(s,\tau_0) ds\right)^2}. \label{dist}
\end{align}
\nn Here $\tau_0$ and $\mathcal{T}_1$ are $t_0$ and $T_1$ with time nondimensionalized by the period of motion instead of $\sqrt{L/g}$, so $\tau_0 = t_0/T$ and $\mathcal{T}_1 = T_1/T$, respectively. With large $\tau_0$ and $\mathcal{T}_1$ we approximate the steady-state long-time average, but here we use moderate values---$\tau_0$ = 3 and $\mathcal{T}_1 = 2$---for computational efficiency. We find that the average speed changes by less than 3\% when $\tau_0$ and $\mathcal{T}_1$ are increased to 5 and 4 respectively. Using $\mathcal{T}_1$ that is not very large also allows for the possibility of
efficient locomotion with nonzero rotation over a period. This can occur as long as the body does not rotate much over time $\mathcal{T}_1$, and we will show that this holds for the optimal motions that we find.

A typical measure of efficiency for sliding locomotion
\cite{HuNiScSh2009a,HuSh2012a,JiAl2013,AlbenSnake2013} is 
\begin{align}
\lambda = \frac{\| \langle \overline{\partial_t \mathbf{X}} \rangle \|
}{\langle P \rangle}, \label{lambda}
\end{align}
\nn the ratio of the average speed to the average power
\begin{align}
\langle P \rangle = \frac{1}{T_1}\int_{t_0}^{t_0 + T_1} \int_0^1  -\mathbf{f}(s,t) \cdot \partial_t\mathbf{X}(s,t) \,ds \,dt = \frac{1}{T_1}\int_{\tau_0}^{\tau_0 + \mathcal{T}_1} \int_0^1  -\mathbf{f}(s,\tau) \cdot \partial_\tau\mathbf{X}(s,\tau) \,ds \,d\tau.
\end{align}
\nn Both $\langle P \rangle$ and $\| \langle \overline{\partial_t \mathbf{X}} \rangle \|$ scale linearly with velocity, but their ratio
$\lambda$ does not. Nonetheless, the solutions depend on the period of motion since it sets the magnitudes of the left-hand sides (inertia terms) in (\ref{fxa})--(\ref{torquea}), so $\lambda$ does vary with the period of the body kinematics $(\Delta\theta_1(t), \Delta\theta_2(t))$, though not as a simple scaling law.

The upper bound on efficiency is
\begin{align}
\lambda_{ub} = \frac{1}{\mbox{min}(\mu_f, \mu_b, \mu_n)}, \label{lambdaub}
\end{align}
\nn corresponding to uniform motion in the direction of least friction, and can be approached by a sequence of particular concertina-like motions, as shown in \cite{alben2019efficient}.
In this work we take the relative efficiency $\lambda/\lambda_{ub}$ as the primary measure of performance. 

\section{Efficient single-frequency (elliptical) kinematics \label{sec:ellipse}}

\begin{figure} [h]
           \begin{center}
           \begin{tabular}{c}
               \includegraphics[width=6in]{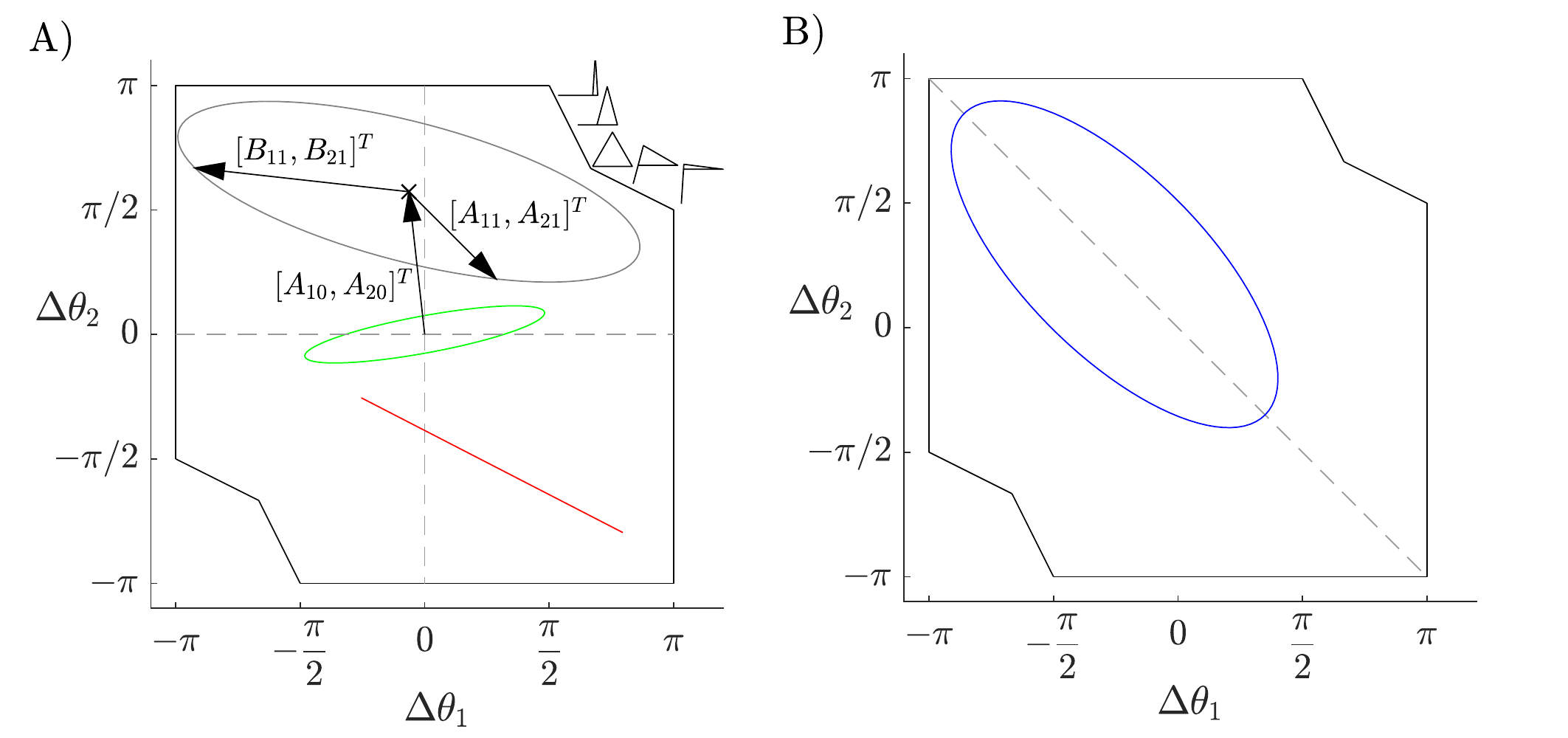} \\
           \vspace{-.25in} \hspace{-.25in}
           \end{tabular}
          \caption{\footnotesize A) Examples of elliptical trajectories in the region of 
non-self-intersecting configurations (inside the black polygonal outline). Examples of body 
configurations along the boundary
of the region are shown at the upper right. The gray ellipse has center $A_{10}, A_{20}$ and
shape given by $\{A_{11}, A_{21},B_{11}, B_{21}\}$. 
B) ($\Delta\theta_1(t)$,$\Delta\theta_2(t)$) for a path symmetric about the line
$\Delta\theta_1 = -\Delta\theta_2$.  
 \label{fig:PathsSchematicInertiaEllipses}}
           \end{center}
         \vspace{-.10in}
        \end{figure}

We begin by considering body kinematics with a single frequency, corresponding to elliptical trajectories in the 
($\Delta\theta_1$,$\Delta\theta_2$)-plane:
\begin{align}
\Delta\theta_1(\tau) = A_{10} + A_{11}\cos(2\pi \tau) + B_{11}\sin(2\pi \tau), \quad
\Delta\theta_2(\tau) = A_{20} + A_{21}\cos(2\pi \tau) + B_{21}\sin(2\pi \tau), \quad 0 \leq \tau \leq 1. \label{GenEllipse}
\end{align}
An example is the gray ellipse in figure \ref{fig:PathsSchematicInertiaEllipses}A, with the coefficient values shown as vectors.
In \cite{alben2021efficient}, we paid particular attention to paths that yield no net rotation of the body over one cycle, and found that they correspond to ellipses with a bilateral symmetry (symmetric under reflection in the line $\Delta\theta_1 = -\Delta\theta_2$, e.g. the blue ellipse in panel B), or antipodal symmetry, i.e. symmetry with respect to reflection in the origin, such as the green ellipse in panel A. With nonzero inertia, such paths may yield nonzero net rotation of the body, but we find nonetheless that the optimally efficient motions are often close to such paths.
A third special case is reciprocal kinematics---degenerate ellipses that reduce to straight line segments, e.g. the red line in panel A. We will see that these can yield efficient locomotion if $\mu_b \gg \mu_f$.







We can rewrite equations (\ref{GenEllipse}) as 
\begin{align}
    \Delta\theta_1(t) =A_{10}+\sqrt{A_{11}^2+B_{11}^2}\cos{(2\pi \tau-\phi_1)} \quad , \quad
        \Delta\theta_2(t)=A_{20}+\sqrt{A_{21}^2+B_{21}^2}\cos{(2\pi \tau-\phi_2)}
\end{align}
\nn for some $\phi_1,\phi_2 \in[0,2\pi)$, and it is clear that the extrema of $\Delta\theta_1$ and $\Delta\theta_2$ are $A_{10}\pm\sqrt{A_{11}^2+B_{11}^2}$ and $A_{20}\pm\sqrt{A_{21}^2+B_{21}^2}$, respectively. To avoid self-intersection, the extrema must not exceed $\pi$ in magnitude, so we must have
\begin{align}
   \sqrt{A_{11}^2+B_{11}^2}<\pi-|A_{10}|,\ \sqrt{A_{21}^2+B_{21}^2}<\pi-|A_{20}|.
   \label{coeffconstraint}
\end{align}
We select a random set of Fourier coefficients that are widely distributed in the space given by (\ref{coeffconstraint}) as follows.
We first choose $A_{10}$ and $A_{20}$ from the uniform distribution on $(-\pi,\pi)$. Then, we choose $A_{11}$ and $A_{21}$ from the uniform distributions on 
$(-\pi+|A_{10}|,\pi-|A_{10}|)$ and
$(-\pi+|A_{20}|,\pi-|A_{20}|)$, respectively. Finally, we choose $B_{11}$ and $B_{21}$ from the uniform distributions on the intervals bounded by $\pm\sqrt{(\pi-|A_{10}|)^2-A_{11}^2}$ and $\pm\sqrt{(\pi-|A_{20}|)^2-A_{21}^2}$, respectively, guaranteeing that (\ref{coeffconstraint}) is satisfied. We eliminate the small number of trajectories that satisfy (\ref{coeffconstraint}) but have self-intersecting configurations that lie outside 
the small
indents in the lower left and upper right portions of the boundary in figure \ref{fig:PathsSchematicInertiaEllipses}A and B.
The only remaining parameter to choose is $R=1/\mu_fT^2$; it is chosen from a uniform distribution on a logarithmic scale from $10^{-3}$ to $10^2$. A lower bound of 10$^{-3}$ for $R$ is used throughout this work because the iterative solver is less robust at smaller $R$, including 0.

We optimize the relative efficiency $\lambda/\lambda_{ub}$ using a population-based optimization algorithm similar to that in \cite{alben2021efficient}. We create a population of 50 individuals, each consisting of seven parameters---the Fourier coefficients in (\ref{GenEllipse}) and $R$---chosen randomly as just described. We update the population through a sequence of generations, typically at least 200. At each generation, we solve the initial value problem for the dynamics of the 50 individuals, and select those with the top 50\% of relative efficiency values. We form two copies of each individual, and add random perturbations to their parameters, drawn from uniform distributions, to form the population of 50 at the next generation.
Motions with self-intersection are discarded at each generation. The sizes of the perturbations are $O\left(10^{-2}\right)$, and decrease with the generation number $N$ as 1/$N$. The optimization is ended when the relative efficiency increases by less than 0.001 in 20 generations. 

We run the algorithm multiple times at each pair of friction coefficient ratios, with different random initializations of the population. The resulting optima are very similar for the different populations, modulo symmetries such as reflection in the line $\Delta\theta_1 = -\Delta\theta_2$.

\begin{figure} [h]
           \begin{center}
           \begin{tabular}{c}
               \includegraphics[width=6in]{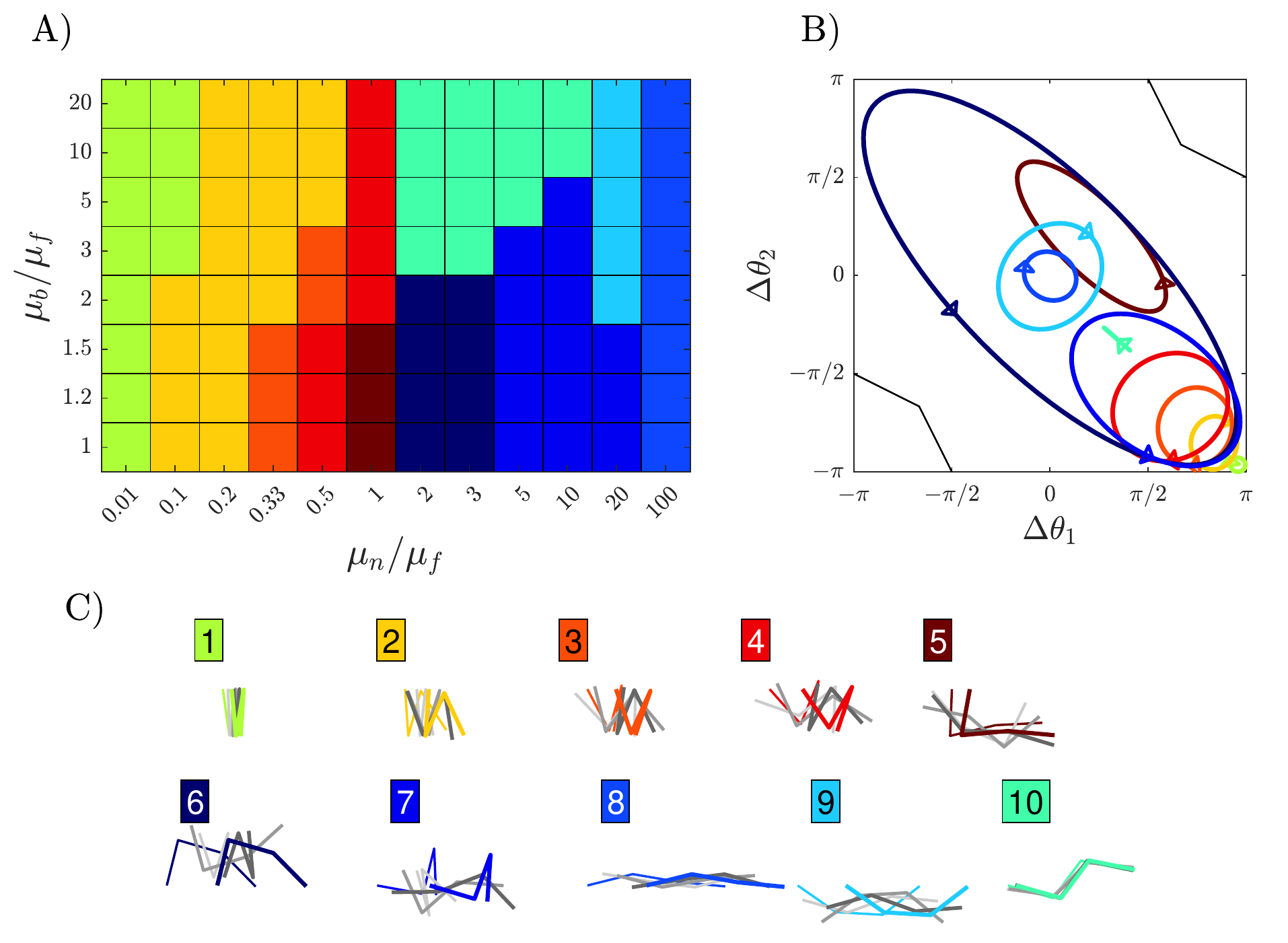} \\
           \vspace{-.25in} \hspace{-.25in}
           \end{tabular}
          \caption{\footnotesize Classification of optimally efficient elliptical trajectories into 10 clusters, each labeled by a distinct color. A) Cluster classification of global optima on a 12-by-8 grid of $(\mu_n/\mu_f,\mu_b/\mu_f)$ values. 
          The color of each square denotes the cluster to which an optimal motion belongs. B) Examples of elliptical trajectories for each of the 10 clusters with the same color in panel A. The ellipses 
          correspond to the motions labeled using the same color in panels C.
 Arrows represent the direction of motion for each trajectory. C) Motions
 for the ten trajectories in panel B, one per cluster, at the following values of $(\mu_n/\mu_f, \mu_b/\mu_f)$: 1) (0.01,2); 2) (0.2,2); 3) (0.33,1.5); 4) (0.5,1.5); 5) (1,1.2); 6) (2,1.2); 7) (10,1.2); 8) (100,2); 9) (20,5); 10) (3,10). Snapshots of the three-link body are shown at five instants spaced 1/4-period apart, starting with the thin colored line and proceeding from light to dark gray, finishing with the thick colored line.
 \label{fig:CharacteristicEllipses}}
           \end{center}
         \vspace{-.10in}
        \end{figure}

We perform the optimization for friction coefficient ratios $(\mu_n/\mu_f,\mu_b/\mu_f)$ ranging over a 12-by-8 grid with values shown on the axes of figure \ref{fig:CharacteristicEllipses}A. We use a k-means clustering algorithm (described in \cite{alben2021efficient}) to group the 96 optima in 10 clusters, denoted by different colors in figure \ref{fig:CharacteristicEllipses}. The clusters are based on the distances between the Fourier coefficients of the optima, but not $R$. Some optimal motions are insensitive to $R$, so $R$ can vary widely in certain clusters even though the Fourier coefficients do not. Including $R$ in the distance calculation would subdivide such clusters.
Figure \ref{fig:CharacteristicEllipses}B plots an example elliptical trajectory for each of the 10 clusters, with the corresponding colors.
The arrows denote the direction of motion of each elliptical trajectory. Within each cluster, the ellipses are relatively uniform. 

As in the no-inertia case in \cite{alben2021efficient}, the clusters in panel A are contiguous in $(\mu_n/\mu_f,\mu_b/\mu_f)$ space. The characteristic ellipses for different clusters are similar in some cases, so we can classify the optima using a smaller set of seven regions in $(\mu_n/\mu_f,\mu_b/\mu_f)$ space, each with a typical motion:

\begin{enumerate}
    \item $\mu_n/\mu_f\ll1$, represented by optima 1 and 2 (numbered in panel C). These optima are small-amplitude oscillations in the lower right corner of panel B, with $|\Delta\theta_1|$ and $|\Delta\theta_2|$ close to $\pi$, meaning that the body is nearly folded together.
    \item $\mu_n/\mu_f<1$ but not $\ll1$ and $\mu_n/\mu_f=1,\mu_b/\mu_f>1$, represented by optima 3 and 4. These optima are similar to 1 and 2, but have larger oscillations and are less folded.
    \item $\mu_n/\mu_f=1,\mu_b/\mu_f\approx1$, represented by optimum 5. The corresponding ellipse (panel B) is not symmetric with respect to $\Delta\theta_1 = -\Delta\theta_2$, unlike the others. This optimum has an undulatory motion with a moderate amplitude.
    \item $\mu_n/\mu_f>1,\mu_b/\mu_f\geq1$, but not $\gg1$, represented by optimum 6. This optimum has the largest amplitude among the ten, reaching two oppositely folded configurations at the upper left and lower right corners of panel B.
    \item $5\leq \mu_n/\mu_f \leq 20$, with $\mu_b/\mu_f$ not $\gg1$, represented by optimum 7. This optimum resembles a concertina motion, with the body folding together before straightening out in each period.
    \item $\mu_n/\mu_f>1$ but not $\gg1$, $\mu_b/\mu_f\gg1$, represented by optimum 10. This optimum has the smallest amplitude, excluding optimum 1, and corresponds to a body that undergoes very small oscillations about a somewhat folded shape, with very small net displacement per period.
    \item $\mu_n/\mu_f\gg1$, represented by optima 8 and 9. These optima oscillate near the straight configuration and perform small-amplitude undulatory motions.
\end{enumerate}

As $\mu_n/\mu_f$ increases from 0.01 to 1 in figure \ref{fig:CharacteristicEllipses}, the optimal trajectories change progressively to larger paths in the lower right corner of panel B. Increasing $\mu_b/\mu_f$ above 1 in this region gives only a modest shift to the cluster boundaries in panel A, that favors slightly smaller trajectories at a given $\mu_n/\mu_f$. The region $\mu_n/\mu_f \geq 1$ has a very different and more diverse set of optimal trajectories (motions 5--10), which are scattered in panel B but still occupy contiguous regions in panel A. 
A similar cluster analysis was performed in \cite{alben2021efficient} for the case of zero inertia, i.e. $R$ fixed at 0. The optimal ellipses in that case were similar to the ones in figure \ref{fig:CharacteristicEllipses} when $\mu_n/\mu_f > 1$. For $\mu_n/\mu_f \leq 1$, the zero-inertia case had many ellipses centered at the origin, unlike the small ellipses in the lower right corner figure \ref{fig:CharacteristicEllipses}B in this regime.
However, when higher frequencies were added in \cite{alben2021efficient}, the optimal trajectories were instead small triangular paths in the corner, like motion 1
in figure \ref{fig:CharacteristicEllipses}C (and corresponding light green path in panel B). 
In \cite{alben2021efficient} we found that the top two local optima can be close in efficiency but have very different kinematics. Perturbations due to the addition of higher frequencies or the presence of inertia could switch the global optimum from one local optimum to another distant one. The same is true here but only for $\mu_n/\mu_f > 1$. Later we will consider the effect of higher frequencies in the present problem with inertia.


\begin{figure} [h]
           \begin{center}
           \begin{tabular}{c}
               \includegraphics[width=6in]{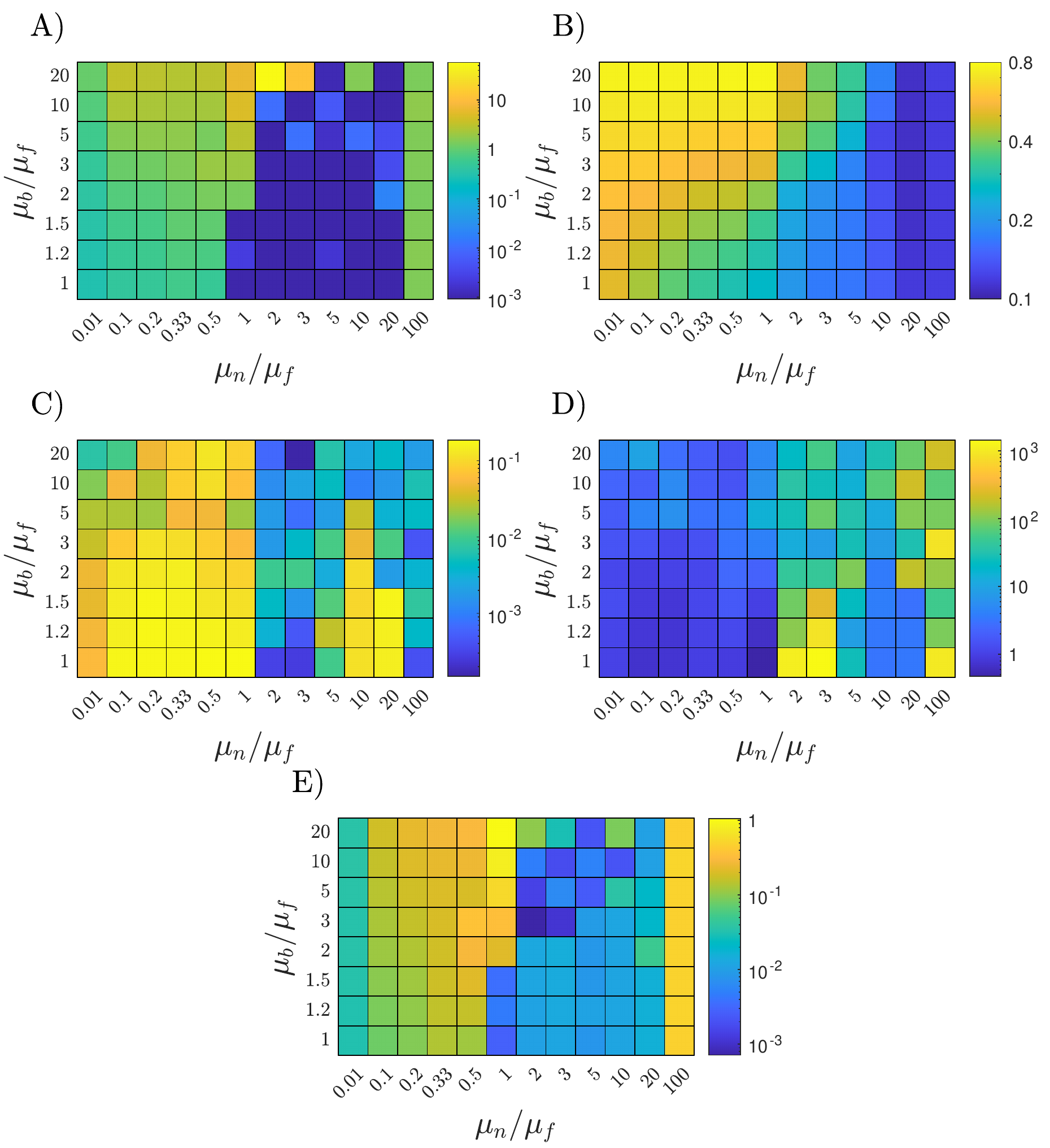} \\
           \vspace{-.25in} \hspace{-.25in}
           \end{tabular}
          \caption{\footnotesize Parameters and performance measures for the single-frequency optima. A) Values of $R=1/\mu_fT^2$. B) Relative efficiencies $\lambda/\lambda_{ub}$. C) Absolute value of net rotation per period (in radians). D) The radius of curvature of the circular path traced by the body's center of mass at times that are integer multiples of a period. E) Average translational speeds (given by equation (\ref{dist})), divided by $\sqrt{\mu_f}$.
 \label{fig:Eff,Rho,MRS}}
           \end{center}
         \vspace{-.10in}
        \end{figure}

The elliptical trajectories in figure
\ref{fig:CharacteristicEllipses}B represent three of the four optimization parameters, the Fourier coefficients but not the value of
$R=1/\mu_fT^2$. The $R$ values are plotted in figure \ref{fig:Eff,Rho,MRS}A, for the
same 96 optima used to construct the clusters in figure \ref{fig:CharacteristicEllipses}A. We can separate $(\mu_n/\mu_f,\mu_b/\mu_f)$-space into four regions based on the optimal value of $R$:

\begin{enumerate}
    \item $\mu_n/\mu_f<1$. $R$ values are moderate $(0.3<R<3)$ here and correspond to optima 1--4 in figure \ref{fig:CharacteristicEllipses}.
    \item $\mu_n/\mu_f\approx1$ and $\mu_b/\mu_f\gg1$. $R$ values are very large: $5<R<60$.
    \item $1<\mu_n/\mu_f\leq 20$ and $\mu_b/\mu_f \leq 10$. $R$ values are very small ($0.001\leq R<0.02)$, often equal to $0.001$, the set lower bound.
    \item $\mu_n/\mu_f=100$. $R$ values are moderate ($1.3<R<1.7)$, and correspond to optimum 8 in figure \ref{fig:CharacteristicEllipses}C.
\end{enumerate}

The most striking feature of figure \ref{fig:Eff,Rho,MRS}A is the sharp jumps from O($10^{-3}$) to O(1) values. Most cases fit the rule that nonzero $R$ is favored at small $\mu_n/\mu_f$ and $R \approx 0$ is favored at large $\mu_n/\mu_f$, but regions 2 and 4 do not. Region 4 coincides with a particular motion, essentially lateral undulation with small-amplitude deflections. When $R$ in region 4 is changed from the optimal value $\approx 1.6$ to 0, the efficiency decreases by 15\%, showing the relative insensitivity of lateral undulation to $R$. When $\mu_n/\mu_f$ is decreased to 20, the efficiency of lateral undulation drops sufficiently that a rather different motion (motion 7 in figure \ref{fig:CharacteristicEllipses}C) becomes optimal, when $\mu_b/\mu_f \leq 1.5$. 

The jump from $10^{-3}$ to $\approx 60$ near the top center of the figure is particularly striking. Figure \ref{fig:CharacteristicEllipses}A shows that the cluster of the optimal motion (10) is the same across the jump, so the sequence of body configurations does not change much. The efficiency of this very-small-amplitude motion turns out to be surprisingly insensitive to $R$, changing by less than 2\% when $R$ changes from $10^{-3}$ to 60, and changing by less than 10\% over the range between these values. This accounts for the rapid changes in $R$ in region 2, across the right side of the top row of figure \ref{fig:Eff,Rho,MRS}A. 
Region 1 has much smaller, but still significant changes in the optimal $R$. It rises almost monotonically by factors of 4--7 moving from the bottom to the top of the five leftmost columns of panel A.

The corresponding relative efficiency values are shown in figure \ref{fig:Eff,Rho,MRS}B.
The efficiency values vary more smoothly than $R$, rising almost monotonically from bottom to top and from right to left. The values on the right side ($\mu_n/\mu_f \geq 2$) are about the same as in the inertia-free case studied in \cite{alben2021efficient}.
In that work, a global peak relative efficiency of 0.58 occurred at $\mu_n/\mu_f = 1$ and $\mu_b/\mu_f = 20$, near the top center, with motion 10 (which is optimal in the region just to the right of this square). In the present work, the global peak relative efficiency also occurs at $\mu_n/\mu_f = 1$ and $\mu_b/\mu_f = 20$, but with a
value of 0.78, with motion 4 instead of motion 10, and with $R$ of 6.2.
The optimal efficiencies in the inertia-free case dropped rapidly as $\mu_n/\mu_f$ decreased below 1, to 0.05--0.2 at the lowest three $\mu_n/\mu_f$ values in panel A. When more frequencies were added, the inertia-free optima were much higher, 0.3--0.4. Here, the efficiencies on the left side of panel A are higher still, 0.5--0.76 in the upper left quadrant. The efficiencies become much smaller when $R$ is decreased for the same kinematics. In general, nonzero inertia allows for much greater efficiency when $\mu_n/\mu_f < 1$, and the optimal kinematics here are sensitive to $R$. As in \cite{alben2021efficient}, the peak efficiencies are much lower when 
$\mu_n/\mu_f \gg 1$.

Figure \ref{fig:Eff,Rho,MRS}C shows the net rotation per period for the optimal motions.
With a nonzero time-averaged rotational speed, points on the body will follow a circle in the $x-y$ plane at times that are integer multiples of a period. Therefore the long-time average of the center-of-mass velocity is zero. However, such cases could yield efficient locomotion over finite
time intervals, whose duration scales inversely with the mean rotational speed. The absolute value of the net rotation per period in panel C is bounded by 0.2 radians, so the direction of locomotion does not change much in one period. However, in some cases (e.g. motions 1 and 10 in figure C) the body also does not travel far in one period, so it is useful to also consider the net rotation per period relative to the center-of-mass displacement per period. For small rotations, this ratio turns out to be the curvature of the circle that the center of mass follows at integer multiples of a period. In figure 
\ref{fig:Eff,Rho,MRS}D we plot the reciprocal of this quantity, i.e. the radius of curvature of the circle traced by the center of mass. On the right side ($\mu_n/\mu_f > 1$), radii are usually $\gg 1$, corresponding to nearly straight trajectories. The smallest radii, 0.5--1, occur in the lower left of the panel, where rotations are non-negligible and net displacements per period are small. The smallest value occurs with isotropic friction, similar to case 5 in figure \ref{fig:CharacteristicEllipses}C, and comparing the thin and thick brown lines it can be seen that the net rotation is not very large even in this case. 

It is also useful to consider the speed of locomotion attained by the optimal motions. Some locomotion studies optimize both the speed and efficiency (or input power) simultaneously \cite{saito2002modeling,ariizumi2017dynamic,baysal2020optimally}, and one can obtain a one-dimensional Pareto frontier of optimal motions in speed-efficiency space. Each such optimum has the highest speed among motions with the same efficiency, or the highest efficiency among motions with the same speed. In general, a fast motion is preferable to a slow motion if both have the same efficiency. Figure
\ref{fig:Eff,Rho,MRS}E shows the dimensionless speed of locomotion (divided by $\sqrt{\mu_f}$) for the 96 optima computed here. The speed values are about 0.5 in the rightmost column ($\mu_n/\mu_f = 100$), where the optima have moderate $R$ values. The speeds are much smaller, 0.001--0.1, in the next five columns to the left ($2 \leq \mu_n/\mu_f \leq 20$), closer to 0.001 where $R \approx 0.001$ (panel A), and closer to 0.1 where $R$ is very large at the tops of these columns. These motions (motion 10 in figure \ref{fig:CharacteristicEllipses}C) have small periods $T$ but also small displacements per period, resulting in small but non-negligible average speeds. The speed has a maximum of 1.06 at the top of the $\mu_n/\mu_f = 1$ column, and decreases gradually moving leftward, to 0.04 in the leftmost column. These are the motions that are nearly folded together, with moderate $R$ and speeds that are small but non-negligible, 0.1--0.2 typically. 

\subsection{Fixed-Period Optimization}
\begin{figure} [ht]
           \begin{center}
           \begin{tabular}{c}
               \includegraphics[width=6in]{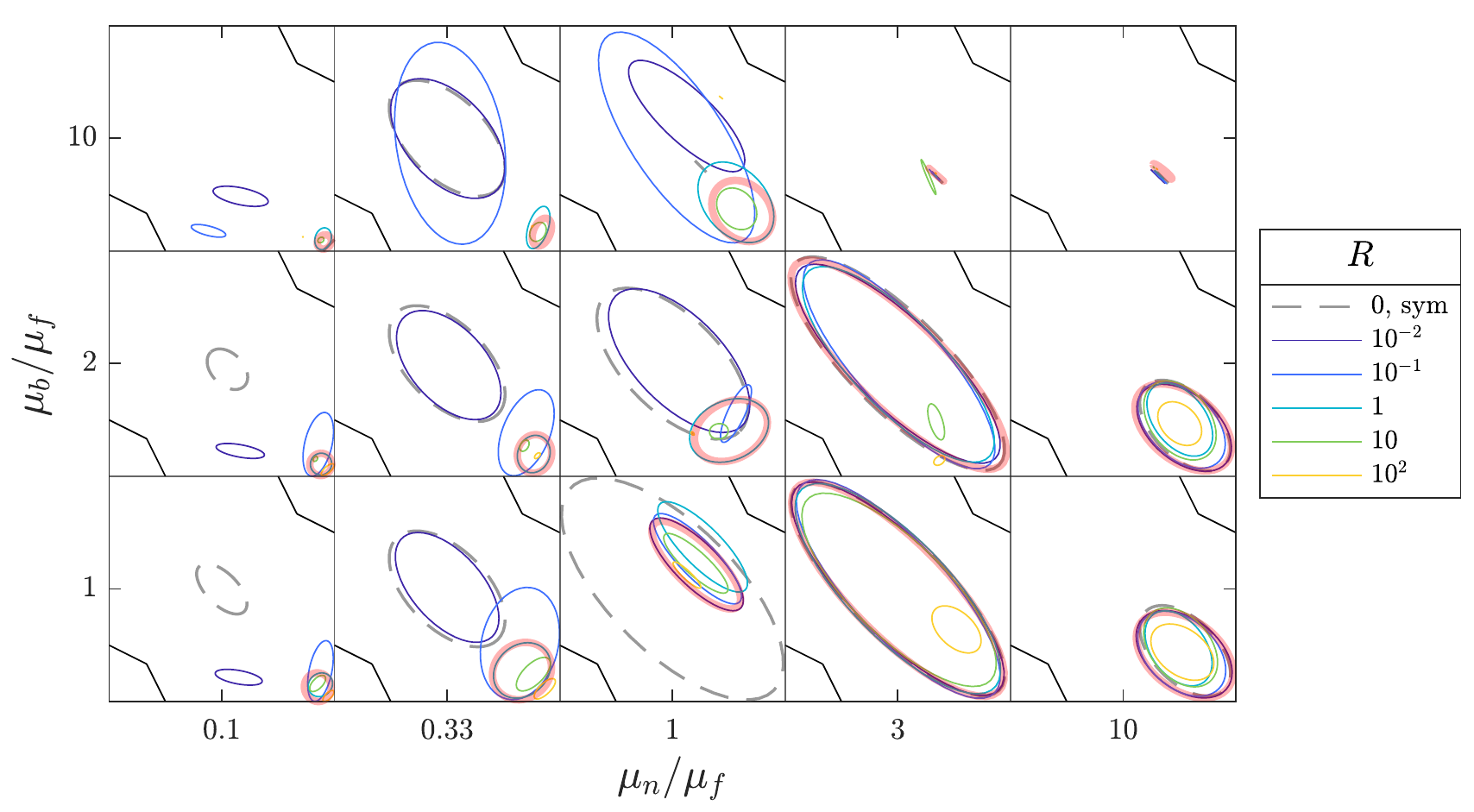} \\
           \vspace{-.25in} \hspace{-.25in}
           \end{tabular}
          \caption{\footnotesize Optimally efficient elliptical trajectories in the $\theta_1$-$\theta_2$ plane for various friction parameters and fixed values of $R=1/\mu_f T^2$, plotted in different colors for different values of $R$. The semi-transparent red ellipses overlaid on this plot represent the optimally efficient elliptical trajectories when $R$ is allowed to vary freely.
 \label{fig:Fixed Period}}
           \end{center}
         \vspace{-.10in}
        \end{figure}
       
We have mentioned that for different optimal motions, the efficiency may have different sensitivities to changes in $R$. We now study more directly how $R$ affects the optimal motions. In \cite{alben2021efficient} we computed optimal three-link motions with $R = 0$. Now we vary $R$ over five orders of magnitude, from $10^{-2}$ to $10^{2}$,
and compute the optimal single-harmonic motions over a 5-by-3 grid of friction coefficient values, a subset of the previous 12-by-8 grid. We follow the same optimization process used earlier, but keep $R$ fixed and vary only the Fourier coefficients in (\ref{GenEllipse}). Thinking of $R$ as an approximate measure of the speed of a motion, this identifies efficient motions at low and high speeds.

In figure \ref{fig:Fixed Period} we plot the optimal trajectories for five fixed $R$ values, listed at right. For comparison, the optimal trajectory from \cite{alben2021efficient}, with $R = 0$ and bilateral symmetry assumed to enforce no rotation, is shown as a gray dashed-dotted ellipse. The trajectory when $R$ is allowed to vary in $[10^{-3}, 10^2]$ to maximize efficiency, as in figures \ref{fig:CharacteristicEllipses} and
\ref{fig:Eff,Rho,MRS} is shown as a thick red ellipse. 
In the rightmost column, $\mu_n/\mu_f=10$, the trajectories are fairly similar for all $R$, so a single type of motion is efficient at both low and high speeds.
Moving leftward to the $\mu_n/\mu_f=3$ column, the same is true except for $R = 10$ and 100 at some $\mu_b/\mu_f$ values. In the left three columns, $\mu_n/\mu_f \leq 1$, there is somewhat more variation with $R$. Here the red global optimum is generally close to the $R = 1$ ellipse, and the $R = 10$ and 100 ellipses are nearby, but sometimes much smaller. Those for $R$ = 0, 0.1, and 0.01 are often much larger and far away from the global optimum, and sometimes from each other. In this region, the optimal motions can be particularly sensitive to $R$ when $R$ is small. This is particularly true for $\mu_n/\mu_f=0.1$. Here and in the top and bottom rows of the $\mu_n/\mu_f=1$ column, the bilaterally symmetric $R = 0$ case is very different from the $R = 0.01$ case. As shown in \cite{alben2021efficient}, at small $R$ there can be distant local optima with similar efficiency values.
In most cases across friction coefficient space, the optimal trajectories with large $R$ (10 or 100) are small ellipses in the lower right corner. These correspond to small oscillations about a mean shape that is nearly folded together.

\subsection{Isotropic friction}

We have noted that in some cases the $R = 0$ and $0.01$ ellipses in figure \ref{fig:Fixed Period} are far from each other. This may be due to the bilateral symmetry assumed only for the $R = 0$ ellipse, and/or to a sensitivity of the optimum with respect to $R$ at small $R$. One such case is isotropic friction, the bottom square in the middle column. This case also has the feature that the red globally optimal ellipse is shifted from the line of bilateral symmetry, $\Delta\theta_1=-\Delta\theta_2$, unlike the other red ellipses in figure \ref{fig:Fixed Period}. The same can be seen in figure \ref{fig:CharacteristicEllipses}B, where the brown ellipse is the only one of the ten that is not approximately bilaterally symmetric. Interestingly, the brown ellipse and all of the isotropic ellipses in figure \ref{fig:Fixed Period} are approximately symmetric with respect to the line $\Delta\theta_1=\Delta\theta_2$.

\begin{figure} [h]
           \begin{center}
           \begin{tabular}{c}
               \includegraphics[width=6in]{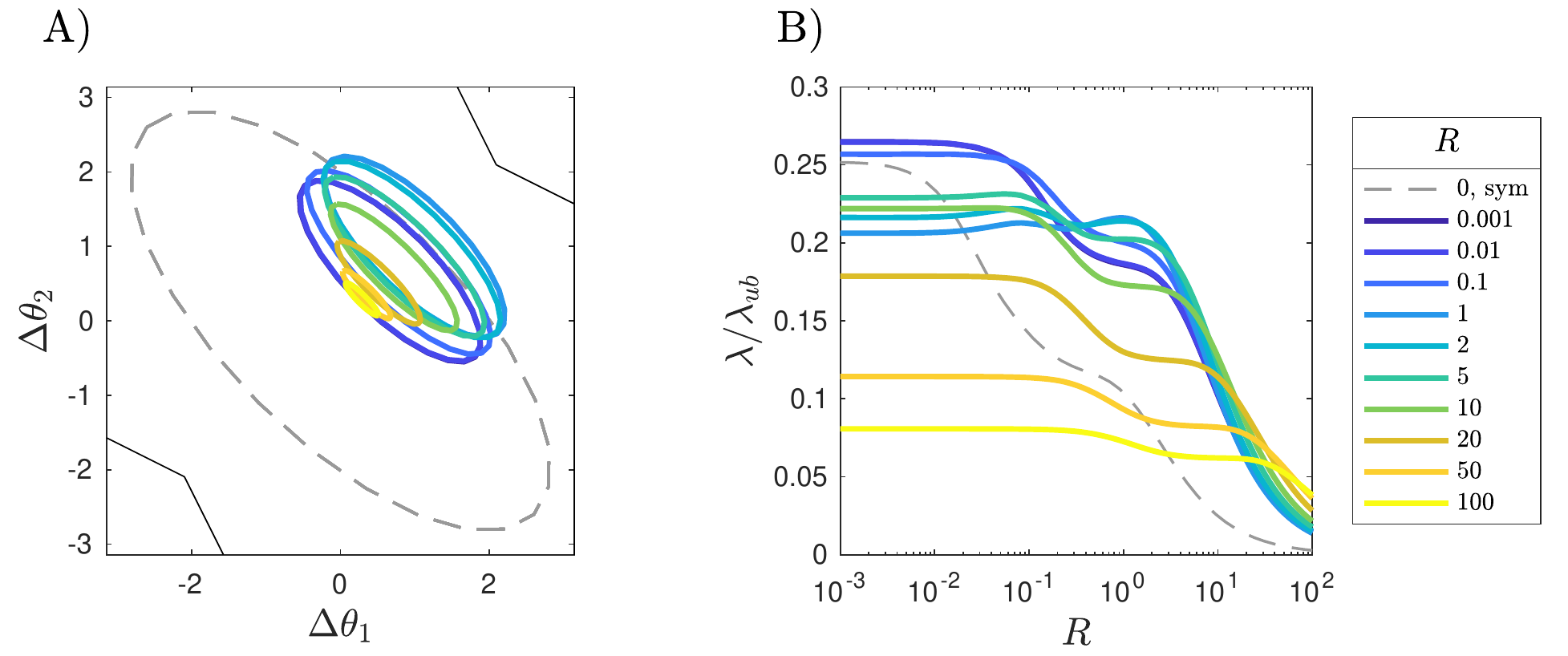} \\
           \vspace{-.25in} \hspace{-.25in}
           \end{tabular}
          \caption{\footnotesize A) Optimal elliptical trajectories for different values of $R=1/\mu_fT^2$ in the isotropic case (when $\mu_n/\mu_f=\mu_b/\mu_f=1$). The plotted ellipses maximize efficiency for each $R\in\{0.001,0.01,0.1,1,2,5,10,20,50,100\}$. The color of each ellipse shows its $R$ value, listed in the legend to the right of panel B. The gray dashed ellipse represents the optimum for the inertia-free bilaterally symmetric case, found in \cite{alben2019efficient}. B) A plot of the efficiency for each ellipse in panel A as $R$ varies. Each line corresponds to the ellipse of the same color.
 \label{fig:IsotropicEllipses}}
           \end{center}
         \vspace{-.10in}
        \end{figure}
We investigate the sensitivity of optimal efficiency with respect to $R$ further in the isotropic case, by examining how efficiencies vary with $R$ for ellipses that are optimal for a given $R$, including those in figure \ref{fig:Fixed Period} and several others.

We set $R$ to each of ten values in the range 0.001--100, perform the optimization, and plot the optimal ellipses in figure \ref{fig:IsotropicEllipses}A (the $R$ values are listed in the legend to the right of panel B). Five of the ellipses are the same as in the isotropic panel of figure \ref{fig:Fixed Period}, as is the gray dashed ellipse, the optimal bilaterally symmetric ellipse with $R = 0$. The ten ellipses with $R \neq 0$ are in the same general region of kinematic space, and appear to change shape continuously with $R$ (the ellipses for $R = 0.001$ and $0.01$ are almost indistinguishable). The most noticeable trend is that the ellipse size decreases monotonically as $R$ increases to 100. With small $R$, i.e. large $T$, the ellipses correspond to low-frequency motions with larger amplitudes, while with large $R$, the ellipses are high-frequency motions with smaller amplitudes. The optimal ellipse with $R = 0.001$ is quite far from the optimal bilaterally symmetric (dashed) ellipse with $R = 0$. Here, removing the constraint of zero rotation allows for slightly more efficient motions that are quite different than those with zero rotation. In \cite{alben2021efficient} with $R = 0$, other non-bilaterally-symmetric ellipses were found with efficiencies close to and even slightly greater than the bilaterally symmetric optimum at certain friction coefficient values.

For each of the eleven ellipses in panel A,
we computed the relative efficiencies for $R$ ranging from 0.001--100, and the values are plotted in panel B.
In each case the efficiencies plateau at small $R$, with the highest efficiency for the $R = 0.001$ ellipse (the graphs for the $R = 0.001$ and $0.01$ ellipses are almost indistinguishable). The efficiency of the bilaterally symmetric ellipse is much more sensitive than the others to changes in $R$ at small $R$, dropping by almost half as $R$ increases from 0.001 to 0.1. At large $R$, the efficiencies of all the ellipses are much smaller, with that of the bilaterally symmetric ellipse well below the others'. Even the ellipses that are optimal at large $R$ have higher efficiency at smaller $R$.
Figure \ref{fig:IsotropicEllipses} shows that in the isotropic case, the family of optimal ellipses with nonzero $R$ vary smoothly but significantly in their shape, location, and performance as $R$ is varied over a wide range.



\section{Multiple-frequency kinematics \label{sec:multiple}}

\begin{figure} [ht]
           \begin{center}
           \begin{tabular}{c}
 \includegraphics[width=6in]{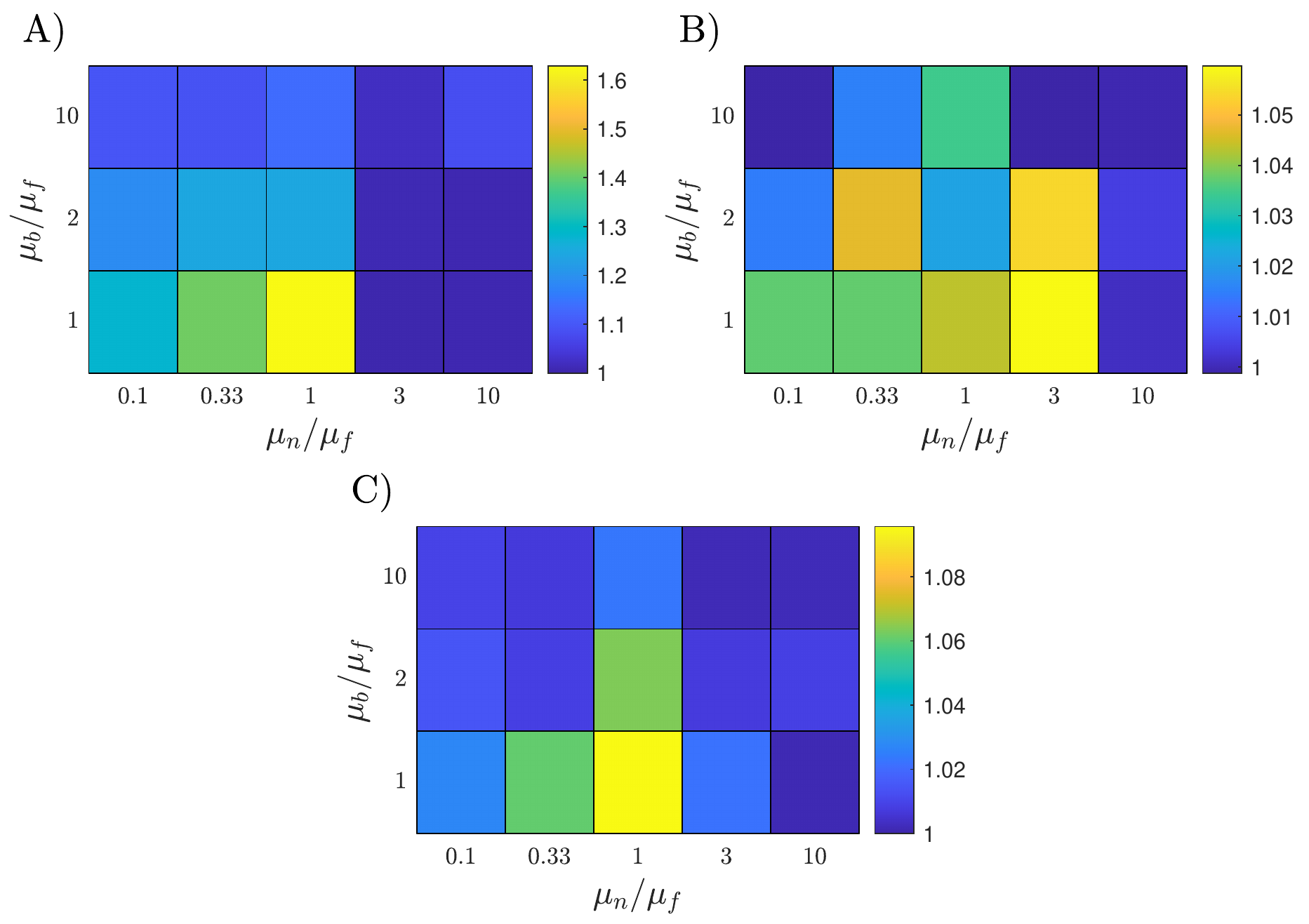} \\
           \vspace{-.25in} \hspace{-.25in}
           \end{tabular}
          \caption{\footnotesize The factor of improvement for the optimal efficiency when the parameter space is enlarged from A) 1 frequency to 2 frequencies; B) 2 frequencies to 3 frequencies; C) 3 frequencies to 4 frequencies.
 \label{fig:AdditionalModeEfficiency}}
           \end{center}
         \vspace{-.10in}
        \end{figure}

We now consider body kinematics with higher frequencies, and check if there are major changes in the optimal motions. With modes up to frequency $n$, (\ref{GenEllipse}) generalizes to
\begin{align}
    \Delta\theta_1(\tau)=A_{10}+\sum_{k=1}^{n}A_{1k}\cos{(2\pi k\tau)}+B_{1k}\sin{(2\pi k\tau)}; \quad
    \Delta\theta_2(\tau)=A_{20}+\sum_{k=1}^{n}A_{2k}\cos{(2\pi k\tau)}+B_{2k}\sin{(2\pi k\tau)}.
\end{align}
We need $4n+2$ parameters to describe a given trajectory up to frequency $n$, so adding higher frequencies enlarges the search space dimension considerably. Also, there is no longer a simple rule
like (\ref{coeffconstraint}) to
ensure that the body does not self-intersect. These factors lead to a large number of invalid and suboptimal trajectories during the optimization. Because of these difficulties, we will analyze the effect of higher frequencies on the smaller 5-by-3 grid of friction coefficient ratios. We have already shown the optima with $n = 1$, and we take small random perturbations of these motions as the
starting population in the search for optima with $n = 2$. We then take the optima with $n = 2$ as the starting point for the $n = 3$ search, and likewise to go from $n = 3$ to 4. We find that this sequential approach to optimization with higher $n$ finds better optima than starting with purely random choices at each $n$. The latter approach could potentially find a wider range of optima, but it tends to stagnate near lower-efficiency motions. 


Figure \ref{fig:AdditionalModeEfficiency} shows the factor of increase in relative efficiency for the optimum with $n = 2$ versus 1 (panel A), 3 versus 2 (panel B), and 4 versus 3 (panel C).
Panel A shows that adding the second frequency gives a significant improvement mainly for $\mu_n/\mu_f\leq1$. For $\mu_n/\mu_f>1$, the improvement factor is less than 1.02 for all cases except for $\mu_n/\mu_f=\mu_b/\mu_f=10$, where it is 1.08. The largest factor of improvement, 1.6, occurs for the isotropic case. 
Moving to panel B, we see that adding a third frequency yields much smaller improvements, by factors of less than 1.06 over the efficiency of the 2-frequency optima, and distributed heterogeneously in friction coefficient space. Adding the fourth frequency (panel C) also gives small improvements, with a distribution similar to that of the second-frequency improvements (panel A).
        
\begin{figure} [ht]
           \begin{center}
           \begin{tabular}{c}
               \includegraphics[width=6in]{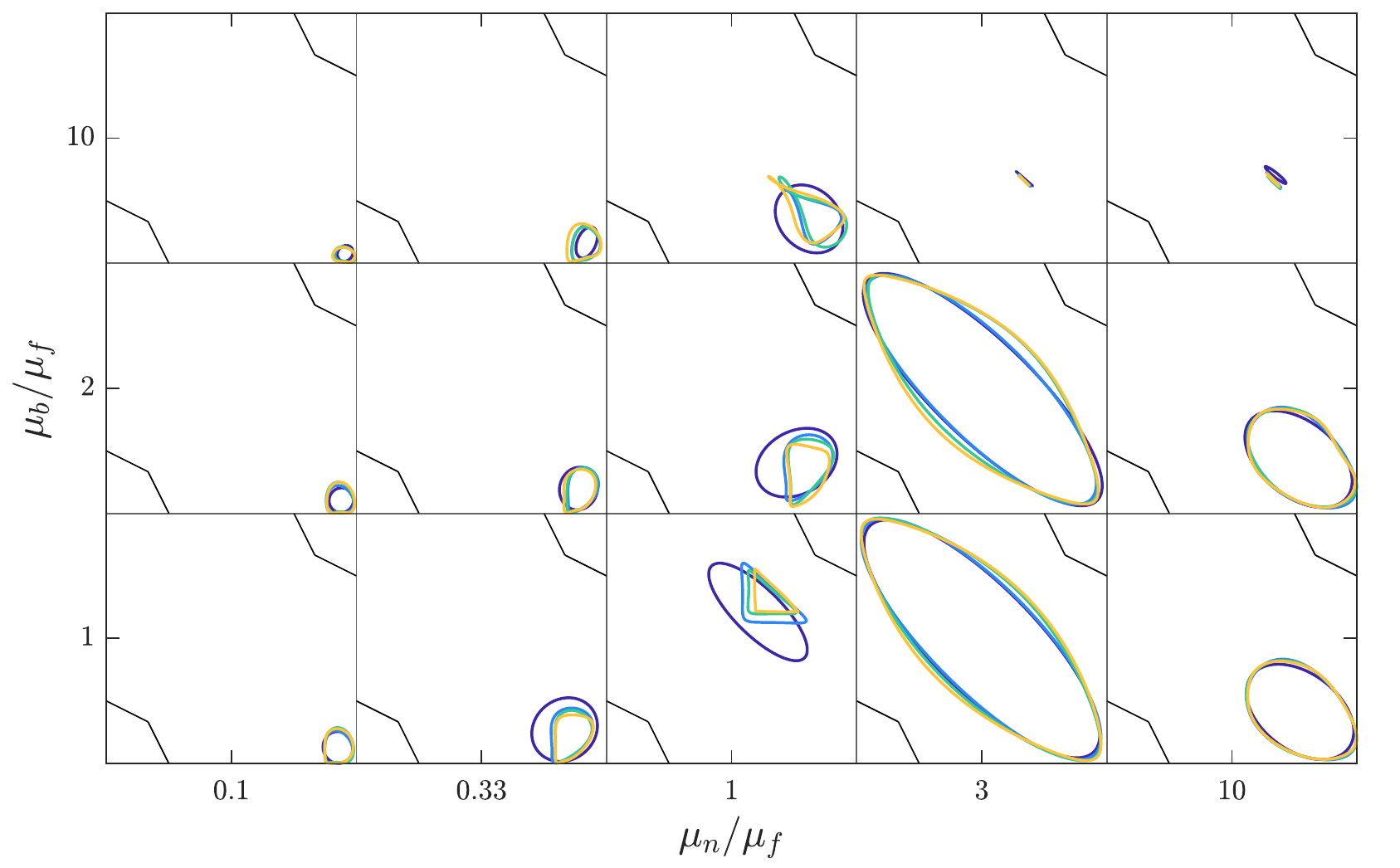} \\
           \vspace{-.25in} \hspace{-.25in}
           \end{tabular}
          \caption{\footnotesize Optimally efficient trajectories with different numbers of frequencies: 1 (dark blue), 2 (light blue), 3 (green), 4 (yellow).
 \label{fig:MultipleModeTrajectories}}
           \end{center}
         \vspace{-.10in}
        \end{figure}
        
Figure \ref{fig:MultipleModeTrajectories} shows the trajectories of the optimal motions in the $\Delta\theta_1$-$\Delta\theta_2$ plane for $n$ = 1--4 frequencies, corresponding to the improvement data in figure \ref{fig:AdditionalModeEfficiency}. In most cases, the single-frequency trajectory is a good approximation to those with higher frequencies, at least in the size and location of the trajectory, with the largest discrepancies around $\mu_n/\mu_f = 1$. The two-frequency trajectories are very good approximations to those with three and four frequencies. The corresponding $R$ values (not shown) also show convergence with the number of frequencies. For the right two columns ($\mu_n/\mu_f \geq 3$) and isotropic friction, $R = 0.001$ for all trajectories. In seven of the eight remaining cases ($\mu_n/\mu_f \leq 1$ and anisotropic), $R = O(1)$ and drops monotonically by 30--55\% going from 1 to 2 frequencies, and then by much less (3--30\%, usually 10--20\%) from 2 to 4 frequencies. In the lone remaining case, the top center panel, $R$ varies nonmonotonically in the range [4,5] for the four trajectories. In general, the small changes with larger $n$ support the possibility that the low-frequency optima here are close to optima that would be obtained as the number of frequencies approaches infinity.

The optimal trajectories with isotropic friction maintain approximate symmetry with respect to the line $\Delta\theta_1 = \Delta\theta_2$ with higher frequencies. The trajectories assume a triangular shape that is very similar to the two-mode optimum with $R = 0$ found by \cite{JiAl2013}. There the friction coefficient ratios, taken from biological snakes, were mildly anisotropic ($\mu_n/\mu_f=1.7$ and $\mu_b/\mu_f=1.3$). The optimal trajectory here with $\mu_n/\mu_f=1$ and $\mu_b/\mu_f=10$ assumes a teardrop shape, maintaining approximate symmetry with respect to the line $\Delta\theta_1 = -\Delta\theta_2$. All the other anisotropic cases are also approximately symmetric with respect to this line, though the deviations from symmetry are noticeable, particularly for the cases that resemble curved triangles near isotropic friction.


\begin{figure} [h]
           \begin{center}
           \begin{tabular}{c}
               \includegraphics[width=5in]{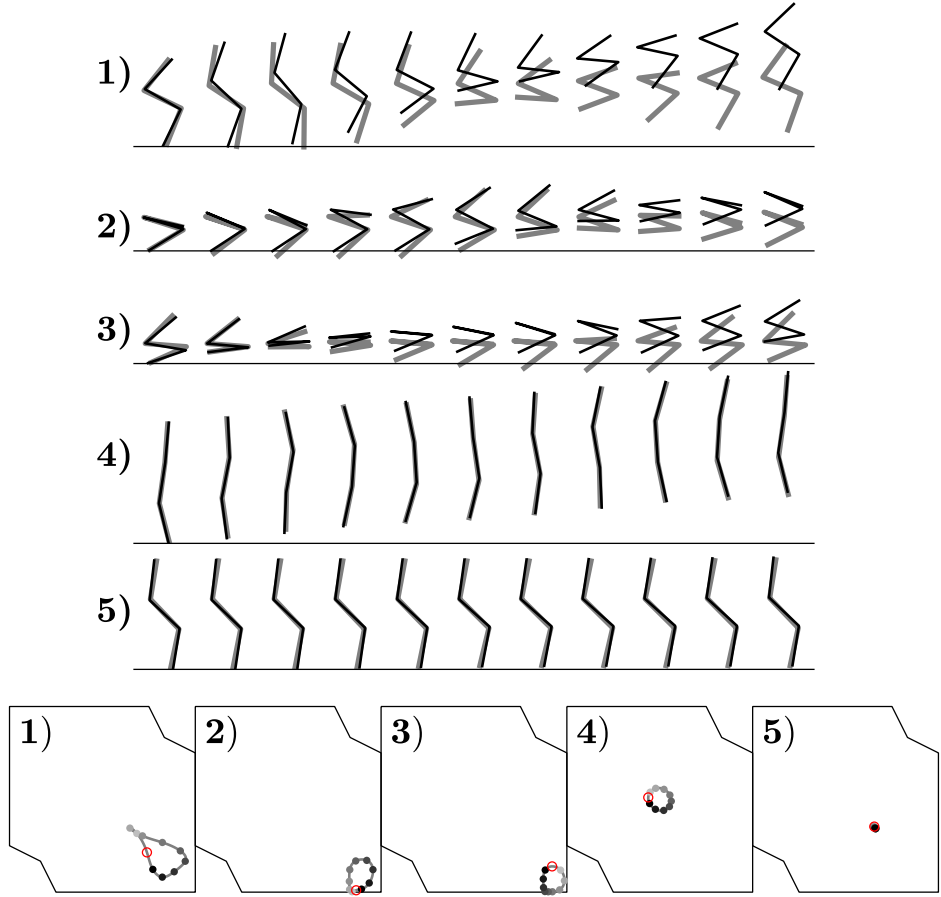} \\
           \vspace{-.25in} \hspace{-.25in}
           \end{tabular}
          \caption{\footnotesize Efficient motions represented by sequences of snapshots over a period. The motions are cases for which moderate inertia ($R = O(1)$) is optimal, and occur at the following values of $(\mu_n/\mu_f,\mu_b/\mu_f)$: 1) $(1,10)$; 2) $(0.33,2)$; 3) $(0.1,1)$; 4) $(100,2)$; and 5) $(2,20)$. 
          The black body represents the optimal motion, and the gray body represents the motion with the same kinematics but no inertia. The snapshots are arranged so the net displacement is vertical, and are spaced uniformly in the horizontal direction from left to right with increasing time. The motions have four frequencies.
 \label{fig:Horizontally spaced trajectories}}
           \end{center}
         \vspace{-.10in}
        \end{figure}
        

We now give examples of the optimal body motions, by plotting sequences of body snapshots over a period of motion. The top five rows of figure \ref{fig:Horizontally spaced trajectories} each show a sequence of snapshots at eleven instants, spaced by 1/10 of a period, from left to right. In each case, the optimal body motion has been uniformly rotated (for ease of comparison) so the net displacement of the body's center of mass over the period is directed vertically up the page. For the actual motion, the net horizontal displacement is zero, but an artificial horizontal displacement has been applied so each snapshot can be seen distinctly from the others. The artificial displacement makes it difficult to see the true horizontal motion, but since it is zero on average, it is less significant than the vertical motion. 

Each row shows two sets of snapshots. The black snapshots are the optimal kinematics (with the optimal $R$ value) at a given friction-coefficient-ratio pair, listed in the caption. The corresponding trajectories in link-angle space are shown at the bottom of the figure, with markers (starting at the red circle and proceeding from light gray to black dots) for the sequence of snapshots. These five cases are chosen to be widely spaced over the regions of friction coefficient space where the optimal $R \geq O(1)$. Figure \ref{fig:Eff,Rho,MRS}A shows the region for the single-frequency case; here the optimal motions have four frequencies, but the relevant friction-coefficient regions are similar. The gray snapshots in each row have the same Fourier coefficients for the link angles as the black snapshots, and therefore the shapes are the same, but $R$ is taken close to zero (0.001). The black horizontal lines mark the starting vertical position of the tail for each set of snapshots, allowing the net displacements for the black and gray snapshots to be seen by comparing the tail positions in the last snapshots to the horizontal line. 
In cases 1--3, the gray snapshots have much smaller net locomotion than the black snapshots, so $R$ has a strong effect on locomotion. Case 4 is an example, already discussed with figure \ref{fig:Eff,Rho,MRS}A,
of lateral undulation for which $R$ has a small effect on locomotion. In case 5,
the difference in displacement is small in absolute terms but significant (27\%) in relative terms.

Comparing the black and gray snapshots in motion 1, we see a noticeable separation between the black and gray bodies during the first half of the cycle, when the bodies straighten, and the black body translates forward slightly more. Here (at the cusp in the link-angle trajectory, below) the body shape is almost fixed, and with inertia the black body can coast forward, which is not possible for the gray body, without inertia. The cusp in the link-angle trajectory is a main difference between the higher-frequency optima and the single-frequency optimum in this case. The rate of separation between the bodies is much larger in the second half of the cycle. Here the front link angle closes and the rear link angle opens. The middle link of the black body moves forward rapidly, while that for the gray body remains almost static. 

Motions 2 and 3 are examples from the large region of optima with $\mu_n/\mu_f < 1$ and $R = O(1)$ (0.24 and 0.48 respectively). The lower panels show that the kinematics are quite similar, but with a phase shift of about half a cycle. In both cases, the rate of separation between the black and gray bodies is largest when the front link angle closes and the rear link angle opens. This happens near the end of motion 2 and near the middle of motion 3. At this moment, the rear link of the black body is almost static and the front and middle links move forward, while for the gray body, the front and middle links are static and the rear link moves backwards, almost canceling out the forward progress on the rest of the cycle.

Motion 4 is an undulatory motion with large $\mu_n/\mu_f$ that is effective both at $R$ = 0.78 (black body) and 0.001 (gray body), but about 18\% more efficient at the larger $R$. For motion 5, the difference in efficiency is smaller, about 4\%, though the difference in $R$ values is much greater (117 versus 0.001).


\begin{figure}
    \centering
    \includegraphics{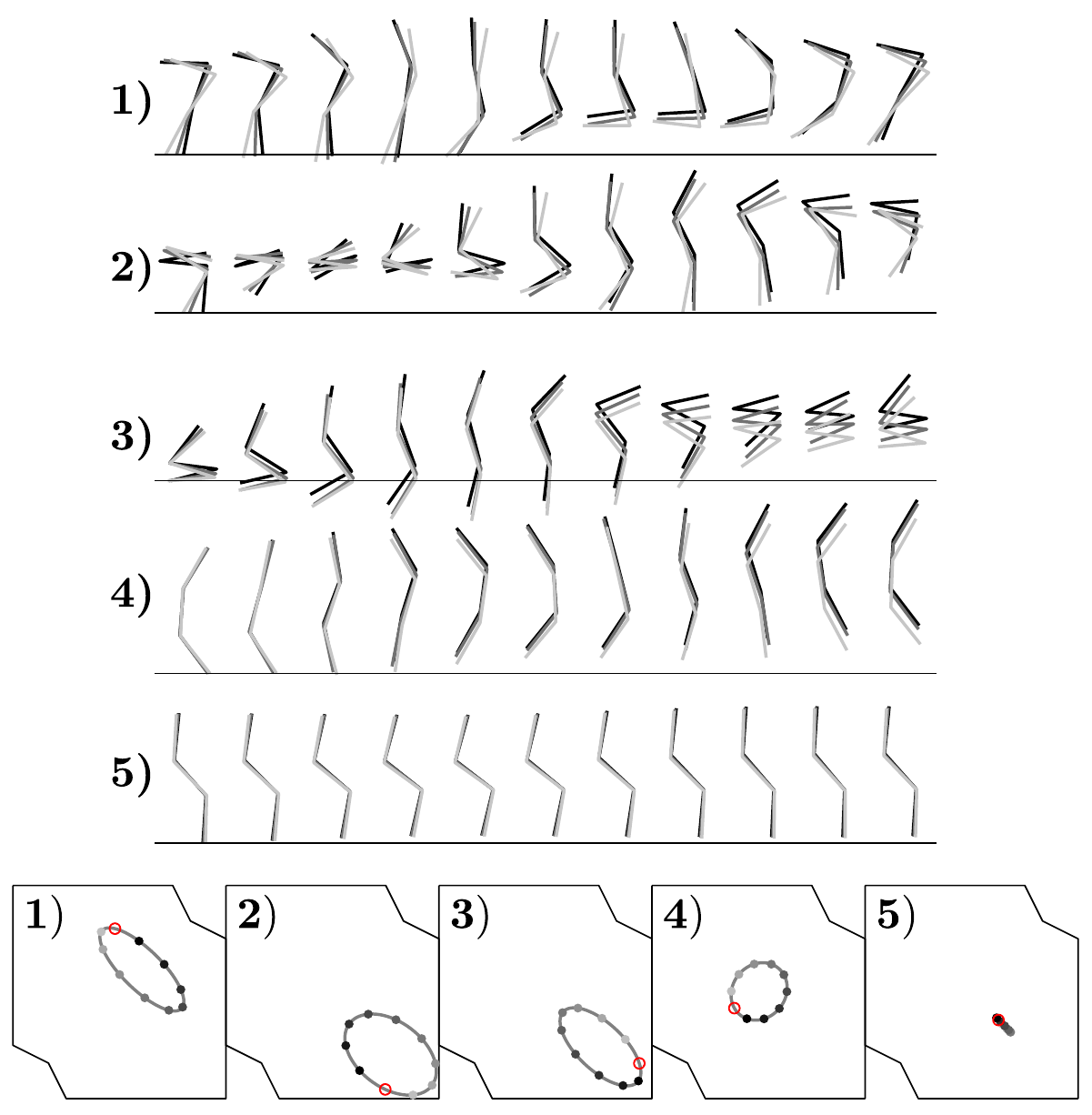}
    \caption{\footnotesize 
    Examples of optimally efficient single-frequency motions that have zero inertia (zero $R$), presented as sequences of snapshots as in figure \ref{fig:Horizontally spaced trajectories}.
    Examples are given at the following values of $(\mu_n/\mu_f,\mu_b/\mu_f)$: 1) $(1,1)$; 2) $(10,1)$; 3) $(10,5)$; 4) $(20,3)$; and 5) $(2,5)$. In each case, three motions are compared. The black body has $R = 0.001$ (optimal), and the medium and light gray bodies have the same kinematics but $R=1$ and $10$ respectively. The snapshots are arranged so the net displacement is vertical, and are spaced uniformly in the horizontal direction from left to right with increasing time.}
    \label{fig: Horiz Elliptical}
\end{figure}

Figure \ref{fig: Horiz Elliptical} shows the same comparison for five cases in the region where small $R$ (0.001) is optimal, the dark blue region in figure \ref{fig:Eff,Rho,MRS}A. Here we use single-frequency optima, which give a good approximation of those with multiple frequencies, but are easier to compute. The motion with the optimal $R$ (0.001 here) is again given by the black snapshots, and is again compared with the motions that have the same kinematics but now with $R$ = 1 (medium gray) and 10 (light gray), to give a sense of the difference that $R = O(1)$ makes. In each case, the relative efficiency and the net displacement per period drop monotonically as $R$ increases from 0.001 to 1 to 10. The net displacements do not vary as greatly as in figure \ref{fig:Horizontally spaced trajectories}; those for $R = 10$ are 45--85\% of those for $R = 0.001$ across the five cases. The relative efficiencies for $R = 10$ are somewhat lower, about 40--75\% of those for $R = 0.001$. Also, the rate of separation between the bodies with small and large $R$ is more uniform across the period than in 
figure \ref{fig:Horizontally spaced trajectories}.
In all five cases, the efficiencies and net displacements undergo another substantial decrease when $R$ is increased from 10 to 100 (not shown).

\section{Summary and conclusions \label{sec:summary}}

This paper has studied the role of body inertia in the efficient sliding locomotion of three-link bodies. Such bodies have been used extensively to model locomotion in viscous fluids and granular media and on dry surfaces. With two spatial degrees of freedom (the angles between the links) and 1--4 temporal frequencies, the number of degrees of freedom is small enough to facilitate optimization methods, but large enough to represent a wide range of motions and to approximate common motions such as lateral undulation. 

We first computed the optima with a single frequency across friction coefficient space. The optimal motions neatly partition the space into a small number of clusters, with variations of one overall behavior---oscillating about a folded-up state---when $\mu_n/\mu_f < 1$, and roughly four different locomotor behaviors, one of which is lateral undulation, when $\mu_n/\mu_f \geq 1$.
The optimal motions oscillate with an $O(1)$ period (the corresponding inertia parameter and the time-averaged speeds are $O(1)$) when $\mu_n/\mu_f < 1$. The optimal period is instead very large (the inertia parameter and the time-averaged speeds are generally very small) when $\mu_n/\mu_f > 1$, except for a few cases with $\mu_n/\mu_f$ or $\mu_b/\mu_f$ very large---
lateral undulation with $\mu_n/\mu_f = 100$ and small-amplitude, high-frequency oscillatory motions with $\mu_b/\mu_f = 20$.

The full set of optimal motions (across friction coefficient space) here is generally similar to the set of zero-inertia optima in \cite{alben2021efficient}, but with inertia there are variations in which motions are optimal at which friction coefficients. There is a large increase in efficiency for $\mu_n/\mu_f \leq 1$ when nonzero inertia is allowed. 

Some of the optima, particularly at large $\mu_n/\mu_f$ (lateral undulation) or large $\mu_b/\mu_f$, maintain their high efficiency when the inertia parameter is varied across a wide range. To understand the effect of the inertia parameter more directly, we computed optima with various fixed values of the inertia parameter. In several cases with small $\mu_n/\mu_f$ we found large changes in the optimal kinematics when the inertia parameter is near zero. With isotropic friction, unlike other friction values, all the optima were strongly asymmetric with respect to 
the line $\Delta\theta_1 = -\Delta\theta_2$ but were symmetric with respect to
$\Delta\theta_1 = \Delta\theta_2$, similar to a zero-inertia optimum with moderately anisotropic friction in \cite{JiAl2013}.

When we progressively increased the numbers of allowed frequencies in the optima motions from from 1 to 2, there were moderate changes, and then much smaller changes in the optimal motions going from 2 to 4 frequencies, indicating that these optima may be close to those with the full Fourier series representation.

Finally, we showed examples of optimal motions with small and with $O(1)$ values of the inertia parameter, and compared them to the motions with the same kinematics but with the inertia parameter in the other regime ($O(1)$ and small, respectively). When the inertia parameter was decreased from an optimal $O(1)$ value to zero for optima with $\mu_n/\mu_f < 1$, the distance traveled (and efficiency) dropped dramatically. At key moments during the motion, the front link moved forward for the optimal motion, while the rear link (and center of mass) moved backward for that with zero inertia. Less dramatic changes were seen with the optima that occurred with zero inertia. The distance traveled and efficiency dropped modestly as $R$ increased from 0.001 to 1 and then to 10, and then more substantially as $R$ increased to 100.

\begin{acknowledgments}
This research was supported by the NSF Mathematical Biology program under
award number DMS-1811889.
\end{acknowledgments}


\bibliographystyle{unsrt}
\bibliography{snake}

\end{document}